\documentclass[fleqn,usenatbib]{mnras}

\usepackage{newtxtext,newtxmath}

\usepackage[T1]{fontenc}

\DeclareRobustCommand{\VAN}[3]{#2}
\let\VANthebibliography\thebibliography
\def\thebibliography{\DeclareRobustCommand{\VAN}[3]{##3}\VANthebibliography}

\usepackage{graphicx}	
\usepackage{amsmath}	
\usepackage[dvipsnames]{xcolor}

\title[V3890\,Sgr in soft X-rays during 2019 outburst]{{\it AstroSat} Soft X-ray observations of the symbiotic recurrent nova V3890\,Sgr during its 2019 outburst}

\author[K. P. Singh et al.]{
K. P. Singh$^{1}$\thanks{E-mail: kpsinghx52@gmail.com (KPS)},
V. Girish$^{2}$,
M. Pavana$^{3,4}$,
Jan-Uwe Ness$^{5}$,
G. C. Anupama$^{3}$
and M. Orio$^{6,7}$
\\
$^{1}$Indian institute of Science Education and Research Mohali, Sector 81, SAS Nagar, Manauli PO, 140306, India\\
$^{2}$Indian Space Research Organisation HQ, Antariksh Bhavan, New BEL Road, Bengaluru 560094, India\\
$^{3}$Indian Institute of Astrophysics, Koramangala, Bengaluru 560034, India\\
$^{4}$Pondicherry University, Puducherry, 605014, India\\
$^{5}$XMM-Newton Science Operations Center, ESAC, Camino Bajo del Castillo s/n, Urb. Villafranca del Castillo, E-28692 Villanueva de la Cañada, Madrid, Spain\\
$^{6}$INAF-Osservatorio di Padova, Vicolo Osservatorio 5, 35122 Padova, Italy\\
$^{7}$Dept. of Astronomy, University of Wisconsin, 475 N, Charter Str., Madison, WI 53706, USA\\
}

\date{Accepted XXX. Received YYY; in original form ZZZ}

\pubyear{2020}

\begin{document}
\label{firstpage}
\pagerange{\pageref{firstpage}--\pageref{lastpage}}
\maketitle

\begin{abstract}
Two long $AstroSat$ Soft X-ray Telescope observations were taken of the third recorded outburst of the Symbiotic Recurrent Nova, V3890\,Sgr.  The first observing run, 8.1-9.9 days after the outburst, initially showed a stable intensity level with a hard X-ray spectrum that we attribute to shocks between the nova ejecta and the pre-existing stellar companion. On day 8.57, the first, weak, signs appeared of Super Soft Source (SSS) emission powered by residual burning on the surface of the White Dwarf. The SSS emission was observed to be highly variable on time scales of hours. After day 8.9, the SSS component was more stable and brighter. In the second observing run, on days 15.9-19.6 after the outburst, the SSS component was even brighter but still highly variable. The SSS emission was observed to fade significantly during days 16.8-17.8 followed by re-brightening. Meanwhile the shock component was stable leading to increase in hardness ratio during the period of fading. $AstroSat$ and {\it XMM-Newton} observations have been used to study the spectral properties of V3890\,Sgr to draw quantitative conclusions even if their drawback is model-dependence.  We used the {\sc xspec} to fit spectral models of plasma emission, and the best fits are consistent with the elemental abundances being lower during the second observing run compared to the first for spectra $\geq$1\,keV. The SSS emission is well fit by non-local thermal equilibrium model atmosphere used for white dwarfs. The resulting spectral parameters, however, are subject to systematic uncertainties such as completeness of atomic data.

\end{abstract}




\section{Introduction}
Novae are binary star systems wherein a white dwarf (WD) accretes matter from a donor that can be a main-sequence star or a red giant, as in some cases. The accreted material forms an accretion disc around the WD. Over a period of time, as the accreted material increases in mass, the temperature at the base of the accretion disc increases until explosive thermonuclear burning of the hydrogen rich material ensues. The base of the accreted layer being electron degenerate, leads to a thermonuclear runaway (TNR) condition and an eruption or an outburst that is observed as a nova \citep[see][for a review]{Starr2012}. The outburst also leads to ejection of the accreted material at velocities $\ge 300$~km~s$^{-1}$  \citep[e.g.][]{GCA2012}. Novae with more than one recorded outburst are referred to as Recurrent Novae (RNe). 
 
The critical mass needed to start a TNR is lower for a high mass WD with high surface gravity. Therefore, RNe are expected to occur on WDs that are near the Chandrasekhar mass limit and accreting at a high rate making them good candidates for being the progenitors of Type Ia supernova. This is one of the reasons for an intensive interest in the study of RNe. 
Readers are referred to comprehensive reviews of RNe by \citet{Sch2010}, \citet{Ness2012}, \citet{GCA2013}, \citet{Muk2014}, and \citet{Or2015}.

Presently there are only 10 confirmed RNe known in our Galaxy, making them very rare events \citep[]{Sch2010,GCA2013}. Four of these are in binaries that contain a red giant mass donor, similar to symbiotic stars, and are sometimes also referred to as symbiotic RNe  \citep*{Sch2010}. These are RS Oph, T CrB, V745 Sco, and V3890\,Sgr. They all have an orbital period of order 1-2 years \citep[see][]{Sch2009}. V3890\,Sgr has an orbital period of 519.6 days and recurrence time of $\sim$ 28 years. The distance of V3890\,Sgr has been found to be 4.36$^{+2.64}_{-1.31}$  kpc from the {\it Gaia} DR2 parallax measurements \citep{Ba2018} with statistical error as given at the {\it Gaia} website {\footnote{gaia.ari.uni-heidelberg.de/tap.html}}. The distance measurements to Novae from $Gaia$ have been discussed in detail by \citet{Sch2018} who point out that these measurements can have large errors which will hopefully be improved in a later release of data.   

The reddening measurement towards the nova obtained from the interstellar features in the optical spectra observed by \citet{Mun19}, leads to $E(B-V)=0.56$, which when compared with interstellar reddening maps indicates a distance of $>$4.5 kpc, which is within the uncertainty range of the {\it Gaia} distance. Here, we assume a distance of 4.4 kpc for V3890\,Sgr.

V3890\,Sgr went into eruption on 2019 August 27.87 \citep*{P2019}, and was immediately confirmed by \citet{St2019} through optical spectroscopy that showed very broad P Cygni profiles of H I and He I lines. Its two previous recorded outbursts were in 1962 \citep*{D1987} and in 1990 \citep*{J1990, L1990}. Based on optical photometry, \cite{Soko2019} suggested that the nova peaked on 2019 August 28.1118 at V=7.17 mag. The earliest emergence of X-rays was recorded on 2019 August 28.438 when a bright X-ray source at the position of the Nova was reported by \citet{Soko2019} from observations with {\it Swift} Observatory  \citep{Bur2005}. An optical spectrum of V3890\,Sgr obtained with the 2m Himalayan Chandra Telescope (HCT) on August 29 by \citet{Pa2019} indicated the presence of weak coronal lines such as [Fe XIV] 5303 \AA, [Ar X] 5535 \AA\ and [Fe X] 6374 \AA. Strong coronal lines from the shock heated plasma were detected during its 1990 outburst as well \citep*{A1994}. RNe with red giant secondary are known to exhibit a shock interaction of the nova ejecta with the red giant wind \citep*{A1998} giving rise to strong hard X-ray (and $\gamma$-rays) and radio emission from the shock heated plasma. Radio emission from V3890\,Sgr was detected with the MeerKAT radio telescope (a precursor for the Square Kilometre Array) at 1.28 GHz by \citet{Ny2019} and with the Giant Metrewave Radio Telescope at 1.4 GHz and 610 MHz by \citet{PaM2019}. $\gamma$-ray emission was also detected by the Large Area Telescope (LAT) on the {\it Fermi} Gamma-ray Space Telescope, during 2019 Aug 26--29 \citep*{Bu2019}.

Soft X-rays in symbiotic RNe are believed to arise predominantly from matter shocked and heated by the nova blast wave from the surface of the WD interacting with the wind from the red giant donor. The shock heated ejecta from these novae evolve extremely rapidly during the outbursts and require almost continuous monitoring. In the recent outburst from V3890\,Sgr, \citet{Or2020} reported a rich emission line spectrum in X-rays from their observations in the 1.2-15 Angstrom range with the $Chandra$ ACIS-S camera and High Energy Transmission Gratings (HETG) on 2019 Sep 3 (starting on day 6.4 since the onset of the optical outburst). The high resolution X-ray spectra revealed several prominent features due to transitions in Fe-L and K-shell ions ranging from neon to iron. A couple of days later, \citet{Page2019a} reported a decrease in the absorption in V3890\,Sgr and emergence of super-soft phase around day 8.57 from their observations with {\it Swift} Observatory. The origin of the super-soft source (SSS) is the residual burning on the surface of the WD \citep[e.g.,][]{Kraut2008}. A deep, continuous, {\it XMM-Newton} observation during the SSS phase was described by \cite{Ness2019}. The high-resolution spectrum taken with the Reflection Grating Spectrometer (RGS)  is dominated by several absorption edges and emission lines along with plasma emission and a blackbody continuum with an effective temperature of $6\times10^5$\,K. Several deep absorption lines, most prominently of H-like and He-like nitrogen and oxygen are measured with the RGS. A deep dip in the middle of the RGS observation seems to have brought the overall spectrum back to the pre-SSS status, suggesting a complete obscuration of all SSS emission.

\begin{figure}
\includegraphics[width=\columnwidth]{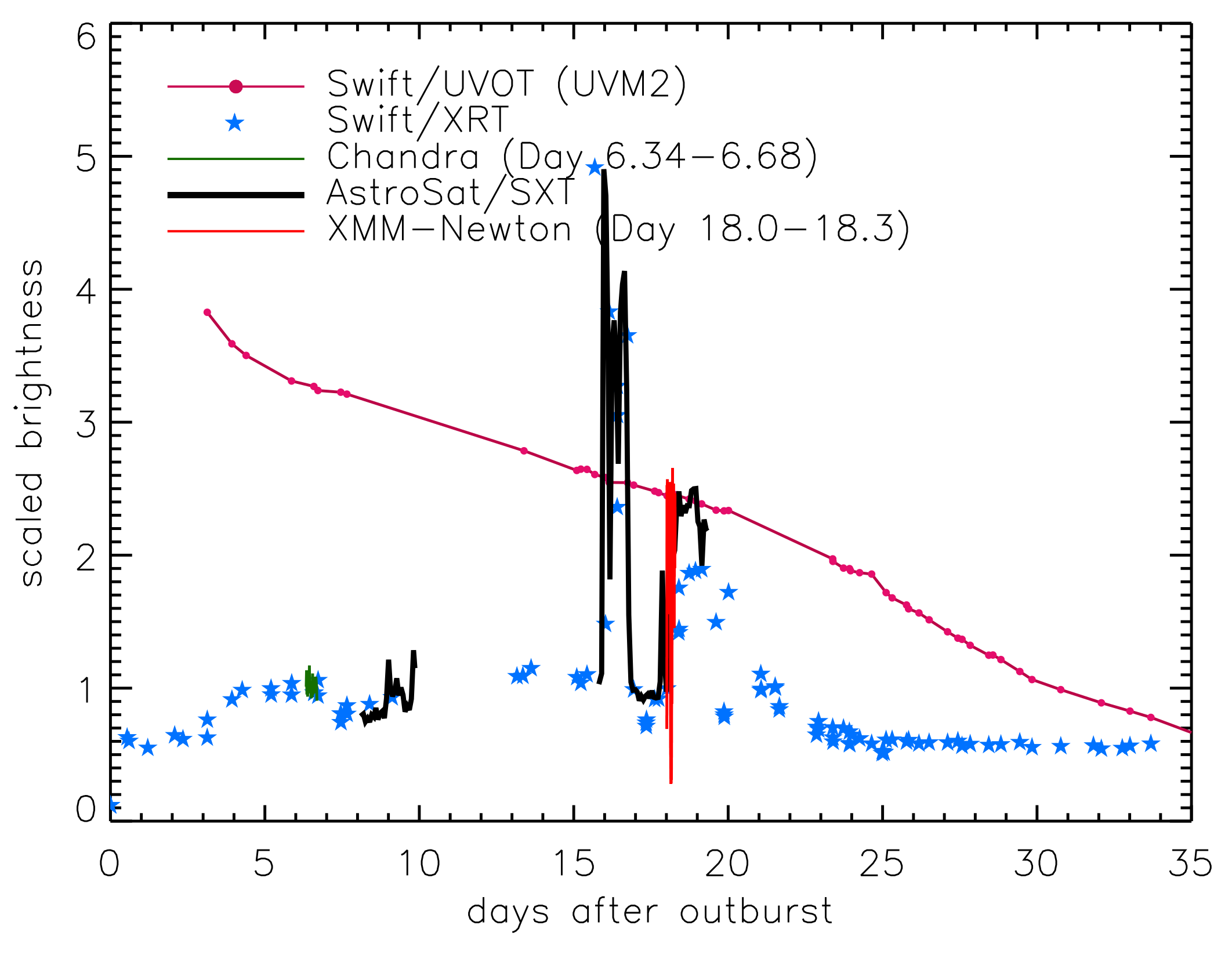}
\caption{The X-ray (0.5 - 10.0 keV) and optical light curve of V3890\,Sgr as obtained with {\it Swift} XRT and UVOT. The $AstroSat$ observations (0.3 - 7.0 keV) are shown superimposed in black. Also shown are the light curves taken with $Chandra$/ACIS (zero order, ObsID 22845) and {\it XMM-Newton}/MOS2 (ObsID 0821560201).}
\label{fig:xrt}
\end{figure}

India's first multi-wavelength space observatory, {\it AstroSat}, \citep{Si2014} observed V3890\,Sgr twice, with its three co-aligned X-ray instruments viz., the Soft X-ray Telescope (SXT), Large Area Xenon Proportional Counters (LAXPC), and Cadmium Zinc Telluride Imager (CZTI). The main instrument used for these observations was SXT \citep{Si2016,Si2017}. The timing and duration of the $AstroSat$ observations are shown in Fig.~\ref{fig:xrt} to put it in perspective with the regular monitoring carried out by the $Swift$. The $AstroSat$ observations were long and carried out with the densest possible monitoring of V3890\,Sgr with a satellite in near-Earth orbit.  The first observation saw the emergence of the SSS phase in V3890\,Sgr accompanied by strong variability \citep*{Si2019a}.  Large amplitude and rapid  variability during the SSS phase was confirmed with $Swift$ observations by \citet{Bear2019}. The SSS phase ended by day 26.2 as reported by \citet{Page2019b}. In the meanwhile, $AstroSat$ SXT had a second long observation of the source during the SSS phase from days 15.9 to 19.6 since the outburst \citet{Si2019b}. The second observation coincided with the peak intensity of the outburst detected with $Swift$ X-ray Telescope (XRT). Dips in the XRT light curve are detected around days 7-8, days 17-18 and day 20.  $AstroSat$ observations provided us with an opportunity to study complex intensity and spectral evolution in soft X-rays -- the quick emergence of the SSS phase, and the rapid variability on time scales of minutes to hours.

In this paper, we present a detailed spectral analysis of X-ray emission observed with the $AstroSat$ SXT on two extended epochs during the recent outburst of V3890\,Sgr, and its spectral evolution. In \S 2 we describe our  SXT observations, and also part of near-simultaneous observation with {\it XMM-Newton}, used for comparison purposes. Section 3 contains the analysis of the X-ray light curves and spectra. Results obtained from fitting non-local thermal equilibrium atmosphere models used for white dwarfs and thin plasma emission models for shocked ejecta are described here.  In \S 4 we discuss our results, compare them with other similar systems, and finally we present our conclusions in \S 5. The results of our monitoring in the optical and radio will be presented in an accompanying paper (M. Pavana et al. in preparation). 


\section{Observations and Data Reduction}
\subsection{AstroSat}
V3890\,Sgr was observed with the $AstroSat$ via a proposal submitted under the Target of Opportunity program  (Observation IDs: 9000003142 and 9000003160, PI: V. Girish) with SXT configured as the prime instrument. V3890\,Sgr was observed throughout an orbit of the satellite while taking care that the Sun avoidance angle was $\ge 45^\circ$ and ram angle (the angle between the payload axis to the velocity vector direction of the spacecraft) > 12$^\circ$ to ensure the safety of the mirrors and the detector.  The first observation (ID 9000003142, hereafter S1) was started on 2019 Sep 5 at 01H:37M:55S UT and ended on 2019 Sep 6 at 20H:19M:27S UT, thus covering the days 8.198 - 9.977 after the outburst. The second observation (ID 9000003160, hereafter S2) was from 2019 Sep 12 at 18H:45M:95S UT to 2019 Sep 16, 07H:14M:32S UT, thus extending from days 15.9 to 19.6 after the outburst  (see Fig.~\ref{fig:xrt} and \S1).  The two observations are not continuous but consist of segments collected during an $AstroSat$ orbit, which are shorter than the orbit due to various observational constraints mentioned above and further data filtering that follows.
\begin{figure}
	\includegraphics[width=\columnwidth]{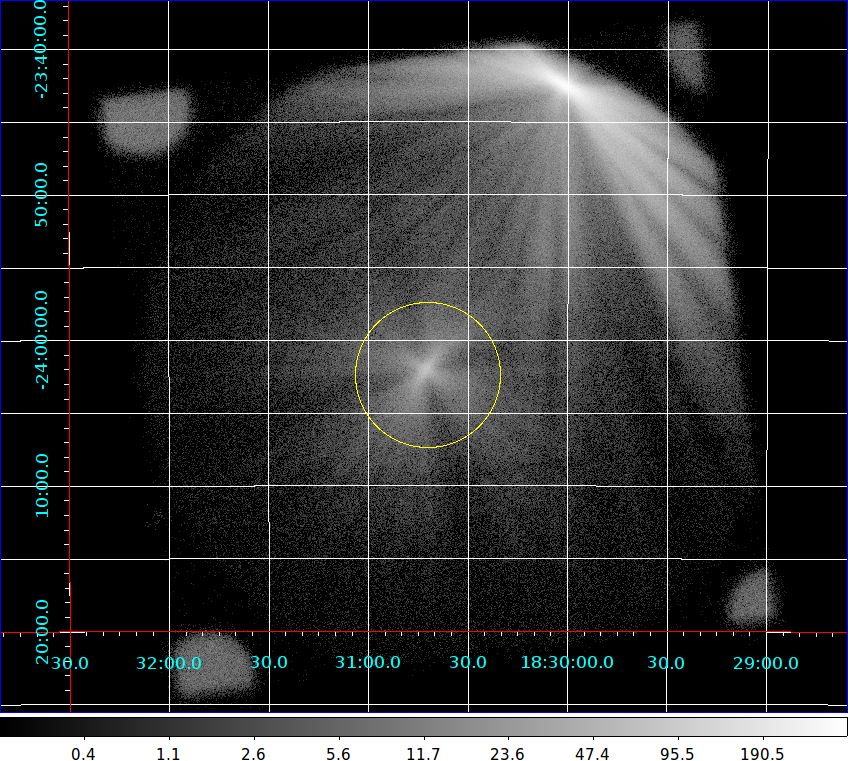}
    \caption{ Soft X-ray image of V3890Sgr in the energy range of 0.3 to 7.0 keV taken during the second epoch of observations with the $AstroSat$ SXT.  Full field of view of the SXT is shown. Log scale is used to show the full dynamic range of brightness. A circle with a radius of 5$\arcmin$ used to extract spectra and light curve is shown. The four bright spots on the corners are the calibration sources.}
    \label{fig:SXT image}
\end{figure}

Data from individual orbits  of the satellite (Level-1 data) were received at the SXT Payload Operation Centre (POC) from the ISSDC (Indian Space Science Data Center). The Level-1 SXT data obtained in the photon counting (PC) mode were first processed with the {\bf $sxtpipeline$} task available as part of the SXT software (AS1SXTLevel2,  version 1.4b) at the SXT POC Website{\footnote{https://www.tifr.res.in/~astrosat\_sxt/sxtpipeline.html}}.
The pipeline calibrates the source events and extracts Level-2 cleaned event files for the individual orbits. The clean process filtered out any contamination by the charged particles due to excursions of the satellite through the  South Atlantic Anomaly region and  occultation by the Earth, and selected only the events with grade 0$-$12 (single-quadrupal events) representing only X-rays and eliminating charged particles \citep*{Bur2005}. The cleaned event files of all the orbits were then also merged into a single cleaned event file using Julia based merger tool developed by G. C. Dewangan to avoid any time-overlapping events from the consecutive orbits. The {\sc xselect} (V2.4d) package built-in {\sc heasoft}{\footnote{https://heasarc.gsfc.nasa.gov/docs/software/lheasoft/}} was used to extract the source spectra and light curves from the processed Level-2 cleaned event files - merged as well as orbit-wise files. 
 Therefore, useful exposure times in each orbit of $AstroSat$ are shorter than its orbit and vary from $\sim$300\,s to $\sim$ 3000\,s.
Total useful exposures obtained are 39894\,s and 65844\,s for the merged data from S1 and S2 observations, respectively. A full field X-ray image from an observation targeting V3890\,Sgr taken with the SXT in the energy band of 0.3-7.0 keV is shown in Fig.~\ref{fig:SXT image}. A bright source, GS\,1826$-$238, which is $21.7\arcmin$ away (towards the top right of Fig. 1) is observed at the very edge of the field of view. Care was taken to minimise the contamination due to scattered X-rays from GS\,1826$-$238 by restricting the extraction of X-ray events from V3890\,Sgr to a circle of radius $5\arcmin$ for light curves and spectra presented in the next section. Since the point spread function (PSF) of images formed in the SXT is large (Full Width Half Maximum of $\sim 2\arcmin$ and Half Power Width of $10\arcmin$) due to scattering by the mirrors and attitude control of the satellite \citep{Si2016}, the extraction is typically carried out for a radius of 10-14$\arcmin$, depending on the brightness of the source. Based on the radial profiles derived for V3890\,Sgr and GS\,1826$-$238 and the PSF measurements \citep{Si2016}, we estimated that in our circle of extraction (Fig.~\ref{fig:SXT image}) the contamination due to GS\,1826$-$238 is $\sim 2-4\%$ for soft X-rays ($<2$~keV) and $\sim 10-12\%$ for hard X-rays ($2-7$\,keV), depending on the intensity of V3890\,Sgr.

\begin{figure*}
	\includegraphics[width=2.0\columnwidth]{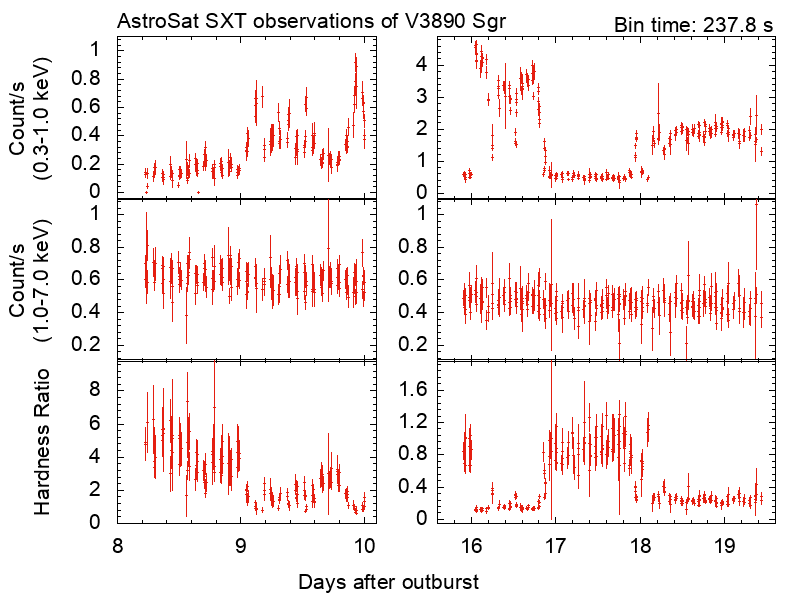}
    \caption{ X-ray light curves and hardness ratios of V3890\,Sgr for both S1 (left panels) and S2 (right panels) observations. Top panels show the light curves in the soft energy band of 0.3 to 1.0 keV. The middle panels show the light curves in the energy range of 1.0-7.0 keV.  The bottom panels show the hardness ratio as described in the text. The X-axis refers to the days after the outburst. Please note that the y-axis scales are different for the top and bottom panels for the S1 and S2 observations. }
    \label{fig:LC1}
\end{figure*}

\subsection{\it XMM-Newton}
{\it XMM-Newton} \citep{Jan01} observed V3890\,Sgr on 2019 September 14, during the second observing run with SXT.  The {\it XMM-Newton} observation under ObsID 0821560201 is part of a larger dataset that is being processed for publication in a forthcoming paper (Ness et al. 2020, in preparation) and here only a subset of the data is used for comparison. We use here the MOS2 observation that was taken in Timing Window mode with the thin optical blocking filter for purposes of comparison with SXT spectra. The observation lasted from 2019 Sep 14 23H:48M:22S UT to 2019 Sep 15 06H:21M:24S UT for useful exposure time of 23.4\,ks. We extracted light curve and spectrum using the {\it XMM-Newton} Science Analysis Software version 17.0.0. For both, light curve and spectrum, we filtered on the same RAWX columns 305-319 in the detector pixel coordinates, and the background was extracted from RAWX columns 260-270. The raw light curve was then corrected with the {\sc epiclccorr} task in SAS. It contains a deep dip shown in Fig.~\ref{fig:xrt} \citep{Ness2019} which we have excluded when extracting the MOS2 (and MOS1 - see below) spectrum by filtering on the time stamp ranges $(6.8489253-6.8490220)\times10^8$\,s and $(6.8490711-6.8491643)\times 10^8$\,s (19\,ks). The pixel coordinates above and the time stamps are given to reproduce the results and have been determined by eye. In addition to source and background spectra, we extracted the detector responses with the {\sc rmfgen} task and the effective area file with {\sc arfgen}. We also extracted the spectrum from the MOS1 detector which was operated in Small Window mode  with thin optical blocking filter. The observation lasted from 2019 Sep 14 23H:48M:06S UT to 2019
Sep 15 06H:25M:33S UT for useful exposure time of 23.6 ks. Owing to the longer readout time, the spectrum is heavily piled up and could only be brought down by excluding the central 250 pixels. The source extraction region was located at (27045,27598) detector pixels with an outer extraction radius of 400 pixels. The background spectrum was extracted from pixel position (28053,26574) with a radius of 180 pixels, somewhat smaller than the source region to avoid source photons to fall into the background extraction region while still fitting into the small window mode.

\section{Analysis and Results}
\begin{figure*}
	\includegraphics[width=2.0\columnwidth]{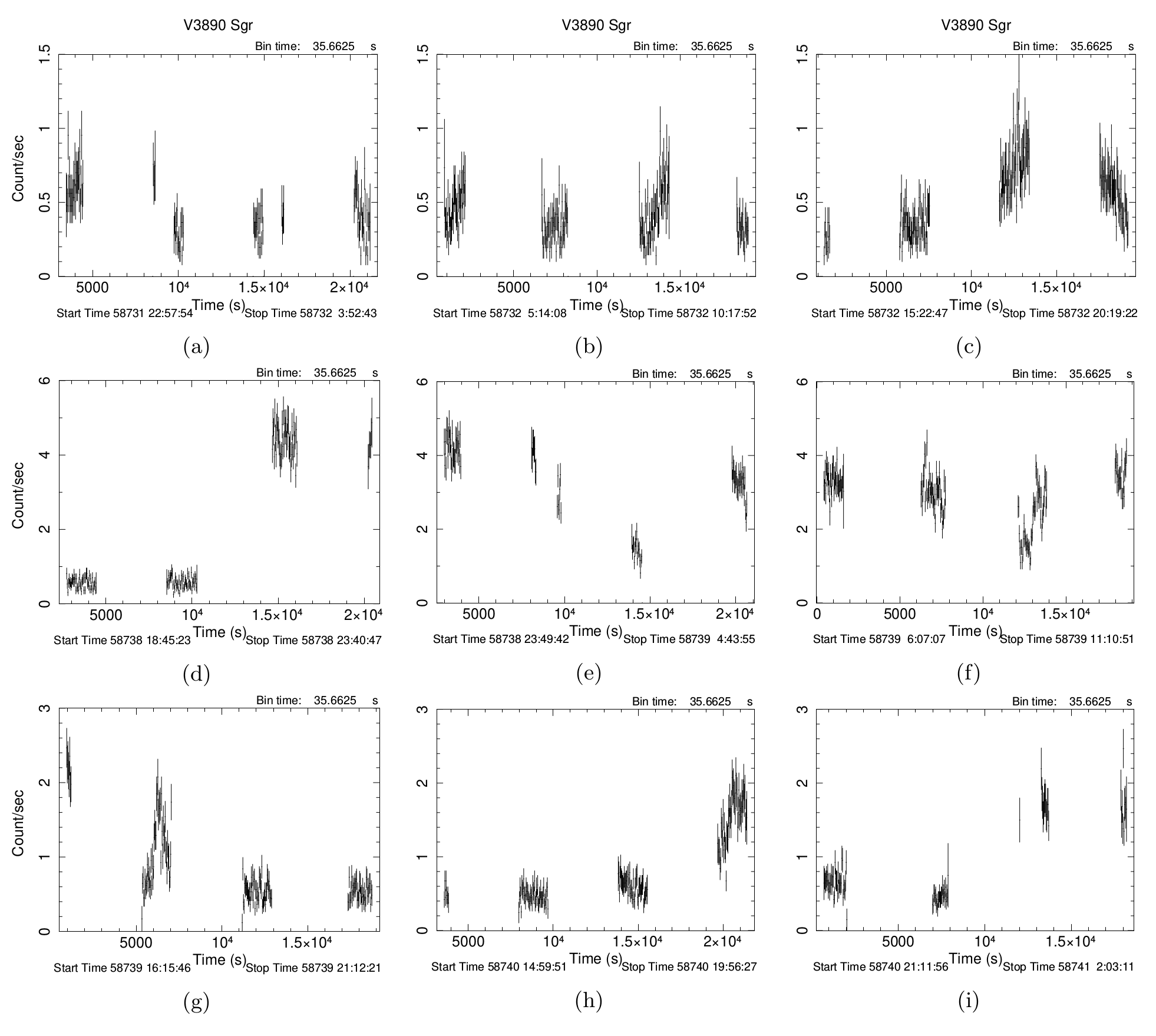}
    \caption {Examples of high time resolution variability in the soft energy band of 0.3-1.0 keV from the S1 observations (panels a to c), and from S2 observations (panels d to i). The times shown here are in units of seconds and the duration shown is in MJD: Mean Julian Day. For reference the start time, t${_0}$, of nova is MJD\,58722.87. The y-axis scale has been chosen to be different to show the range and details of variations, appropriately.}
    \label{fig:Var}
\end{figure*}

\subsection{X-ray light curves}

SXT light curves of V3890\,Sgr extracted from the two observations S1 and S2, are shown in Fig.~\ref{fig:LC1} for two energy bands: soft band of 0.3 $-$ 1.0\,keV (top panels) and hard band of 1.0$-$7.0\,keV (middle panels). The data were binned in time bins of 237.8\,s. These light curves show that a) the count rates in the soft band are highly variable, b) the average count rates during S1 observation are more in the high energy band than in the softer band, c) the intensity in the hard energy band is considerably weakened vis-a-vis the softer energy band in the S2 observations. The overall trends in the hard band count rates are also reflected in the hardness ratios plotted in the bottom panels of Fig.~\ref{fig:LC1}, where the hardness ratio is defined as the ratio of count rates in the hard band divided by the count rates in the soft band. The count rates in the hard band, however, appear to be nearly constant with a count rate of 0.607$\pm$0.004 for the S1, and 0.456$\pm$0.003 for the S2 observations. The variance seen around the constant value is 231.5 for 199 data points in S1 data, and 346.5 for 337 data points. 
Light curves shown are not background subtracted. The same observation cannot be used for extracting the background for the observation for the reasons of PSF mentioned in the previous section. The background in SXT used here is estimated from deep observations of source free regions at mid-Galactic latitudes (32$^{\circ}$ and -32$^{\circ}$) and is found to be very steady at 0.012$\pm$0.005 count s$^{-1}$ in the soft band and 0.014$\pm$0.005 count s$^{-1}$ in the hard band, in this nearly equatorial and low-Earth orbit satellite after the usual screening of events described in previous section and for the same extraction radius as used for V3890\,Sgr. Thus the background is almost negligible here compared to the count rate registered from V3890\,Sgr. 
The hardness ratio variations observed (Fig.~\ref{fig:LC1}) appear to be in anti-correlation with the variations seen in the soft band.  

Short term variability was detected in the soft band (<1\,keV) emission, as already reported in \cite{Si2019a}. These rapid variations were further studied by creating light curves with shorter time bins of 35.6625\,s. Some examples of rapid (minutes to hours time scales) variability are shown in Figure~\ref{fig:Var}. These variations are seen to be mostly sporadic and random and are difficult to characterise with any particular mathematical form or structure, like flares etc.

\begin{figure*}
\includegraphics[width=2.0\columnwidth, angle=0]{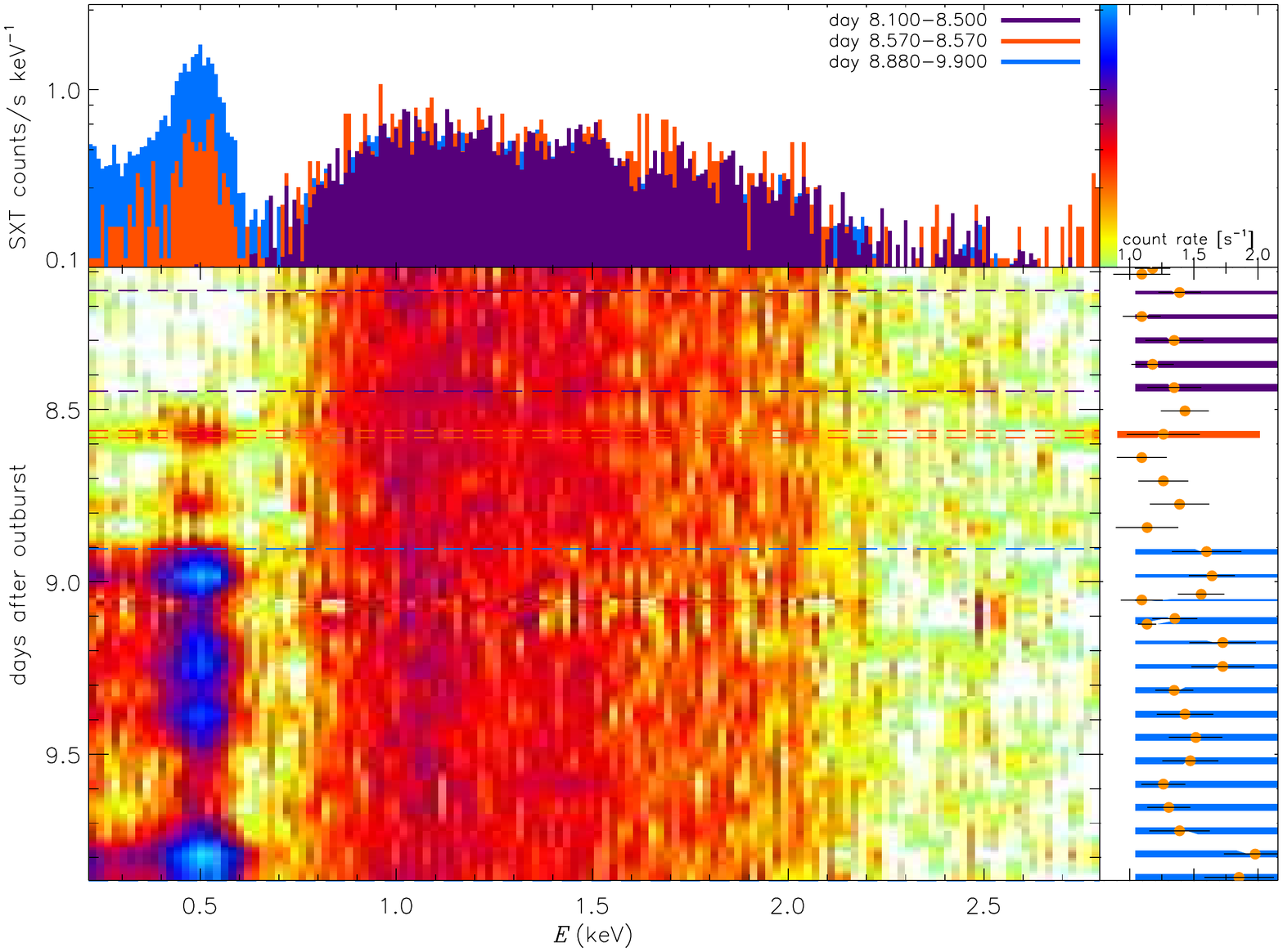}
    \caption{Spectral evolution map from the observation S1 with the $AstroSat$ SXT. The top panel displays spectra extracted from time intervals corresponding to the horizontal dashed lines of the same colours in the time map below and in the light curve to the right (rotated by 90$^\circ$). The central panel shows the spectral time evolution with elapsed observing time along the vertical axis and photon energy on the horizontal axis, on the same scales as the light curve in the right and the spectra on top, respectively. The brightness is colour coded, and for each colour, the corresponding count rate can be derived from the colour bar in the top right.}
    \label{fig:smap1}
\end{figure*}

The light curves were also extracted with the highest available time resolution of 2.3775s in the SXT for period search in the soft band, the hard band and total energy band of 0.3-7.0 keV. The light curves were first subjected to Fast Fourier Transform using the {\it powspec} tool from the {\sc xronos} package in {\sc heasoft} version 6.20. As the SXT light curves are not  continuous but have several gaps due to {\it AstroSat}'s low-Earth orbit, the output power spectra from {\it powspec} has many false peaks. 
To search for real periods, these periodograms were passed through the {\it clean} algorithm by \citep{hoegbom74,roberts87} as implemented in Interactive Data Language (IDL) by \citet{fullerton97} using a low gain of 0.001 and passing through 30,000 iterations. The high energy data ($>$1.0 keV) do not show any significant periodicity. Multiple peaks common to both S1 and S2 observations were detected in the power spectrum of the low energy ($<$1 keV) data corresponding to 0.56d, 1.67h and 1.59h. These seem to be related to the observation duration and orbital period of {\it AstroSat} and, therefore, are not related to the source.

\begin{figure*}
\includegraphics[width=2.0\columnwidth, angle=0]{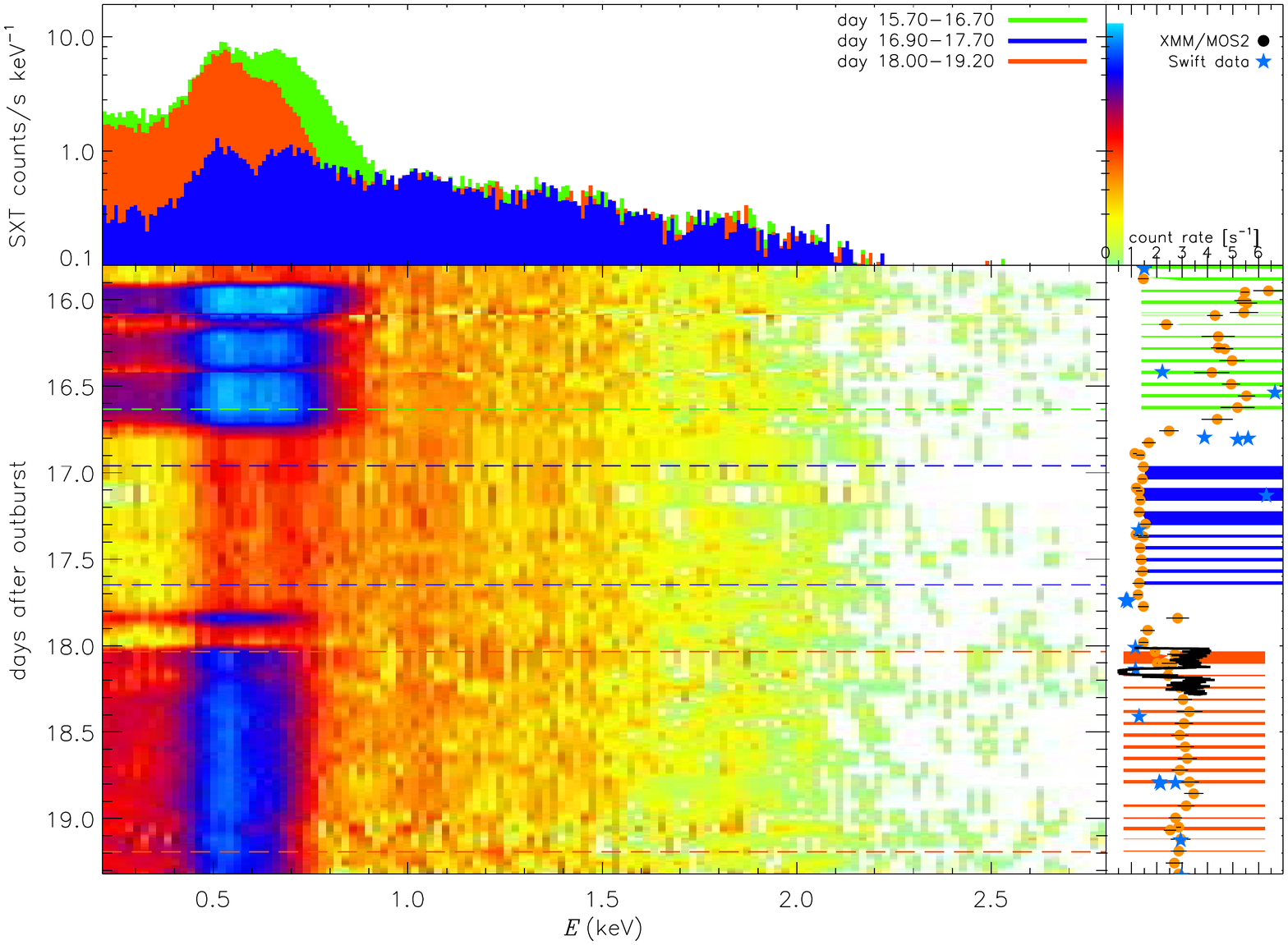}
    \caption{Same as in the Fig.~\ref{fig:smap1} but for the S2 observation with the $AstroSat$ SXT. Also shown on the right panel, at 90$^\circ$, are scaled counts from near simultaneous measurements with $Swift$ XRT (blue stars) and {\it XMM-Newton} (black line, \citealt{Ness2019}).}
    \label{fig:smap2}
\end{figure*}

\subsection{X-ray Spectra} 
\begin{figure*}
	\includegraphics[width=2.0\columnwidth]{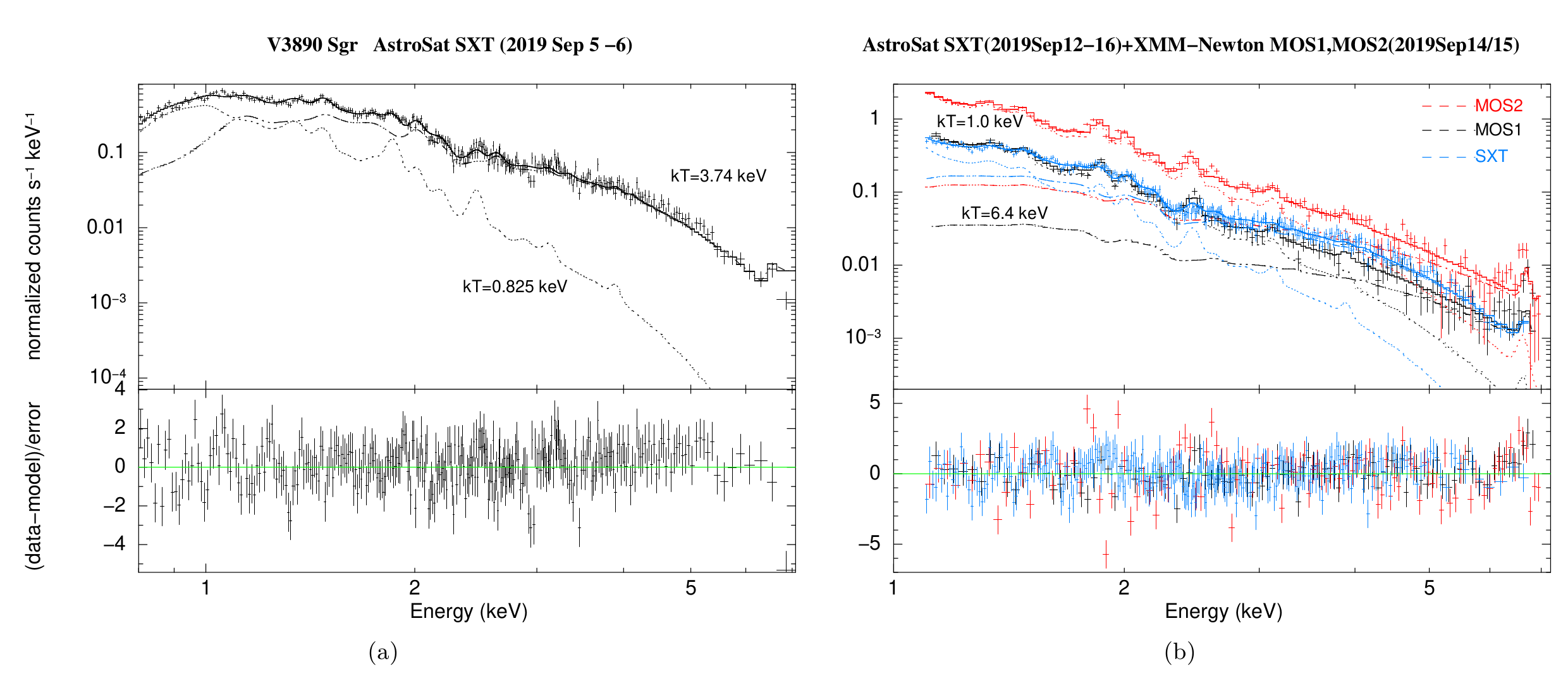}
    \caption{ Merged hard X-ray spectra for S1 observation (left panel) and S2 observation along with the MOS1+MOS2 observation (right panel). The spectra shown are for the energy range 0.8$-$7.0 keV for S1, and 1.1$-$7.0 keV for S2, MOS1 and MOS2, for reasons as explained in the text. Best fit models consisting of two temperature optically-thin plasmas (Model 1 for S1; Model 3 for S2 + MOS1 + MOS2) are shown as histograms. Contribution of each spectral component is also shown. All data points and lines in black show MOS1 data and models, red color show MOS2 data and models, and blue lines show SXT data and models.}
    \label{fig:hard}
\end{figure*}

X-ray spectra were extracted for the duration of
each orbit of $AstroSat$  amounting to 28 spectra in observation S1, and 51
spectra during S2, to understand the variability seen in V3890\,Sgr over
time scales of hours.  All the individual spectra derived from the duration
of each orbit of the satellite for the S1 observation and S2 observation
are shown in Fig.~\ref{fig:smap1} and Fig.~\ref{fig:smap2}, respectively.
These figures reiterate that almost all the spectral variability is
confined to the soft energy bands (as was also depicted in
Fig.~\ref{fig:LC1}). A clear increase in the flux for energies below
0.8\,keV, coincident with the onset of the SSS phase around day 8.57 after
the outburst (see \S1) can be seen in S1 observations shown in
Fig.~\ref{fig:smap1}. This figure also shows that the flux above 0.8\,keV
is almost constant and that the variability is confined to energies below
0.8\,keV, thus marking a natural boundary between the variable low energy
spectral hump and the nearly constant emission above that energy. The
delineation in energy between the variable intensity portion and the
constant intensity portion is not as clear in the case of S2 observation
(Fig.~\ref{fig:smap2}), however.   In the case of S2 observations, the
variable parts of the spectra show a double hump extending in energies to
$\sim$1.1\,keV, and the nearly constant intensity part (3rd hump) begins
only at energies greater than $\sim$1.1\,keV.  Since the flux at higher
energies ($>$0.8\,keV for S1 and $>$1.1\,keV for S2) is nearly constant
during each orbit of the observation, though varying from S1 to S2
observations separated by $\sim$6\,days, we also derived two merged X-ray
spectra $-$ one using all the data from the S1 observations, and another
using all the data from the S2 observations, thus creating two spectra with
long exposures for the study of weak hard X-ray emission part of the
spectrum that hardly varies. The two SXT spectra with long exposure, S1 and
S2, are shown in Fig.~\ref{fig:hard}. Because these hard portions of X-ray
spectra vary from S1 to S2 (see \S3.1), we modelled these X-ray spectra
separately.
The lower limit for the energy used for spectral analysis of
the non-variable component was 0.8\,keV for the S1 observation and 1.1\,keV
for the S2 observation thus avoiding the variable softer component
completely in S1 and to a large extent in S2.  Modelling the steady high
energy part of these two spectra with high signal-to-noise ratio allowed us
to freeze the model parameters for the steady part, thus enabling the study
of the highly variable low energy spectra from each of the orbits (with
short exposures and low signal-to-noise ratio) by varying only a few
parameters (see \S3.2.1 below). This also enabled us to compare SXT data
with contemporaneous $Chandra$ and {\it XMM-Newton} observations (see
below). 

The spectra were modelled using the {\sc xspec} program (version 12.9.1; \citealt{Ar1996}) distributed with the {\sc heasoft} package (version 6.20). A background spectral file ”SkyBkg\string_comb\string_EL3p5\string_Cl\string_Rd16p0\string_v01.pha”, derived from a composite of several deep sky observations of source free regions taken from mid-Galactic latitudes (see \S3.1), and distributed by the instrument team is used in the analysis of all the spectra analysed here. We used the ancillary response file (ARF)
"sxt\string_pc\string_excl00\string_v04\string_20190608.arf".
Similarly the spectral response file used in this work is sxt\string_pc\string_mat\string_g0to12.rmf.
The spectral and ancillary responses, and the background files are available at the SXT POC website (see \S2) {\footnote{ https://www.tifr.res.in/~astrosat\_sxt/index.html}}.
The counts in the spectra were grouped using the $grppha$ tool to ensure a minimum of 25 counts per bin, prior to further analysis here and below. We used $\chi^2$ statistic to assess the goodness of the fit.

Following the modeling of the $Chandra$ HETG spectra by \citet{Or2020}, where the intensity level was about the same as in the case of our S1 observation that was carried out $\sim$2 days after the $Chandra$ observation (see Fig.~\ref{fig:xrt}), the merged hard S1 spectrum (with about the same energy band as in HETG spectra) shown in the left panel of Fig.~\ref{fig:hard} was modelled by fitting a two component thin thermal plasma in collisional ionization equilibrium. This plasma model, known as $bapec$  (Broadened Astrophysical Plasma Emission Code) in {\sc xspec} calculates emission spectrum from collisionally-ionized diffuse gas using the AtomDB atomic database (version 3.0.7{\footnote{$http://www.atomdb.org/index.php$}}) incorporating velocity- and thermally-broadening.  We used the line shifts and line broadening values reported by \citet{Or2020} as fixed parameters, while the temperatures of the two $bapec$ models were kept as free parameters. The specific values of the line blue shifts and thermal velocities were: $z=-0.00289$ and $v=632$ km/s for the low temperature component and $z=-0.00265$ and $v=1572$ km/s for the high temperature component, respectively.  The two plasma emission components were multiplied by a common absorber model, $Tbabs$ (Tuebingen-Boulder absorption model) for interstellar medium with a single free parameter $N_\text{H}$, i.e., the equivalent neutral hydrogen column density in the line of sight to the source. We used the baseline solar elemental abundances as listed in $aspl$ \citep{As2009} in {\sc xspec}. The abundances of all the elements in the plasma component were tied together to vary as a single parameter with respect to the solar values.  The best fit parameters, viz., the column density, the temperatures, the emission measures of the plasma components (and their abundances) were obtained by using the $\chi^2$ minimization technique as used in the {\sc xspec}, and are given in Table~\ref{tab:Tab1} along with their 90\% confidence errors. We first tried the solar abundance values by freezing the common abundance parameter to unity (Model 1 in Table~\ref{tab:Tab1}) and later by making it free (Model 2 in Table~\ref{tab:Tab1}) which led to a slightly improved fit (the minimum reduced $\chi^2$ (henceforth $\chi^2_\nu$) going down to 1.207 for 302 degrees of freedom (DOF) from 1.22 for 303 DOF) to the data (see Table~\ref{tab:Tab1}). The best fit abundances in Model 2 are nearly solar for both the temperature components. The best fit values for two temperatures and $N_\text{H}$ remain unchanged in both the models. The values of $N_\text{H}$ and kT$_1$ are slightly lower than those obtained by \citet{Or2020} from $Chandra$ data, while the kT$_2$  value is about the same considering the error bars. The best fitted  Model 1 and contributions of two spectral components are shown in the left panel of Fig.~\ref{fig:hard} for the S1 spectrum.

\begin{table}
	\centering
	\caption{ Best fit spectral parameters for Two Temperature plasma models with a common multiplicative absorber models fitted to S1 data for energies $>$ 0.8 keV. The errors and upper limits quoted are with 90\% confidence. Blue shift (z) and line width values have been fixed from \citet{Or2020}. Flux values quoted at the end are independent of the models.}
	\label{tab:Tab1}
	\begin{tabular}{l l}
		\hline
		Parameter & Value \\
		\hline
		Model 1: $Tbabs(bapec+bapec)$ &\\
		Solar Abundances & \\
		\hline
		$N_\text{H}$ (cm$^{-2}$) & 10.1 $\pm$ 0.6 $\times$ 10$^{21}$\\
		kT$_1$ (keV)  &  0.825 $\pm$ 0.024\\
		Blue shift z & -0.00289\\
		Line Width (km s$^{-1}$) & 632 \\
		EM$_1$ (cm$^3$) & 5.9 $\times$ 10$^{57}$\\
		kT$_2$ (keV)  &  3.74 $\pm$ 0.21\\
		Blue shift (z) & -0.00265 \\
		Line Width (km s$^{-1}$) & 1572 \\
		EM$_2$ (cm$^3$) & 1.06 $\times$ 10$^{58}$ \\
		 $\chi^2_\nu$/Degrees of freedom (DOF) & 1.22/303\\
		\hline
		Model 2: $Tbabs(bapec+bapec)$ & \\
		Variable Abundances & \\
		\hline
		Z/Z$_{\odot}$  & 0.83 $\pm$ 0.13 \\
		$N_\text{H}$ (cm$^{-2}$) & 9.8 $\pm$ 0.7 $\times$ 10$^{21}$\\
		kT$_1$ (keV)  &  0.83 $^{+0.07}_{-0.02}$\\
		EM$_1$ (cm$^3$) & 6.75 $\times$ 10$^{57}$\\
		kT$_2$ (keV)  &  3.77 $^{+0.25}_{-0.22}$\\
		EM$_2$ (cm$^3$) & 1.10 $\times$ 10$^{58}$ \\
		$\chi^2_\nu$/DOF & 1.207/302\\
		\hline
		Model 3: $Tbabs(bvapec+bvapec)$ & \\
		Variable Abundances from Table 2 of \citet{Or2020} & \\
		\hline
		$N_\text{H}$ (cm$^{-2}$) & 8.11 $\pm$ 0.5 $\times$ 10$^{21}$\\
		kT$_1$ (keV)  &  0.82 $\pm$ 0.22\\
		EM$_1$ (cm$^3$) & 3.5 $\times$ 10$^{55}$\\
		kT$_2$ (keV)  &  3.47$^{+0.17}_{-0.14}$ \\
		EM$_2$ (cm$^3$) & 9.8 $\times$ 10$^{57}$ \\
			 $\chi^2_\nu$/DOF & 1.325/303\\
		Flux$_{0.8-7.1 keV}$ (ergs cm$^{-2}$ s$^{-1}$) & 5.1 $\times$ 10$^{-11}$\\
	\end{tabular}
	\hrule
\end{table}	

\begin{table}
	\centering
	\caption{Best fit spectral parameters for Two Temperature plasma models with a common multiplicative absorber models fitted jointly to S2, MOS1 and MOS2 data for energies $>$ 1.1 keV. The errors and upper limits quoted are with 90\% confidence. The range of EM values reflects the normalisation constants measured with three detectors. Blue shift (z= -0.00289) has been fixed from \citet{Or2020}, and the line width is assumed to be nil. Flux values quoted at the end are independent of the models. A systematic error of 3\% was applied while fitting the data.} 
	\label{tab:Tab2}
	\begin{tabular}{l l}
		\hline
		Parameter & Value \\
		\hline
		Model 1: $Tbabs(bapec+bapec)$ &\\
		Solar Abundances & \\
		\hline
		$N_\text{H}$ (cm$^{-2}$) & 6.5 $\pm$ 0.4 $\times$ 10$^{21}$\\
		kT$_1$ (keV)  &  0.96 $\pm$ 0.02\\
		EM$_1$ (cm$^3$) & (4.3-9.3) $\times$ 10$^{57}$\\
		kT$_2$ (keV)  &  4.9 $^{+0.6}_{-0.5}$ \\
		EM$_2$ (cm$^3$) & (1.2-5.3) $\times$ 10$^{57}$\\
		 $\chi^2_\nu$/Degrees of freedom (DOF) & 1.72/488\\
		\hline
		Model 2: $Tbabs(bapec+bapec)$ & \\
		Variable Abundances & \\
		\hline
		Z/Z$_{\odot}$  & 0.47 $^{+0.060}_{-0.054}$  \\
		$N_\text{H}$ (cm$^{-2}$) & 6.0 $\pm$ 0.5 $\times$ 10$^{21}$\\
		kT$_1$ (keV)  &  0.95 $\pm$ 0.025\\
		EM$_1$ (cm$^3$) & (0.7-1.5) $\times$ 10$^{58}$\\
		kT$_2$ (keV)  &  5.7 $^{+1.2}_{-0.7}$\\
		EM$_2$ (cm$^3$) & (1.1-5.2) $\times$ 10$^{57}$ \\
		$\chi^2_\nu$/DOF & 1.48/487\\
		\hline
		Model 3: $Tbabs(bvapec+bvapec)$ & \\
		Variable Abundances & \\
		\hline
		Ne/Ne$_{\odot}$  & 2.2$^{+1.1}_{-0.9}$ \\
		Mg/Mg$_{\odot}$ & 0.45$^{+0.09}_{-0.08}$\\
		Si/Si$_{\odot}$ & 0.64$^{+0.09}_{-0.08}$\\
		S/S$_{\odot}$ & 0.64$^{+0.12}_{-0.11}$\\
		Fe/Fe$_{\odot}$ & 0.46$^{+0.15}_{-0.13}$\\
		$N_\text{H}$ (cm$^{-2}$) & 5.0 $\pm$ 1.5 $\times$ 10$^{21}$\\
		kT$_1$ (keV)  &  1.0 $\pm$ 0.04\\
		EM$_1$ (cm$^3$) & (0.6-1.1) $\times$ 10$^{58}$\\
		kT$_2$ (keV)  &  6.4$^{+1.8}_{-1.1}$ \\
		EM$_2$ (cm$^3$) & (0.9-4.6) $\times$ 10$^{57}$ \\
				 $\chi^2_\nu$/DOF & 1.40/483\\
		Flux$_{SXT: 1.1-7.1 keV}$ (ergs cm$^{-2}$ s$^{-1}$) & 3.0 $\times$
		10$^{-11}$\\	
		Flux$_{MOS1: 1.1-7.1 keV}$ (ergs cm$^{-2}$ s$^{-1}$) & 2.4 $\times$10$^{-11}$\\
		Flux$_{MOS2: 1.1-7.1 keV}$ (ergs cm$^{-2}$ s$^{-1}$) & 1.6 $\times$10$^{-11}$\\
	\end{tabular}
	\hrule
\end{table}	

We then fixed the elemental abundances individually for Ne, Mg, Al, Si, S, Ar, Ca, and Fe (relative to the solar values) to the values determined by \citet{Or2020} and fitted the S1 spectrum with the $bvapec$ models replacing the $bapec$ models for the two temperature components, i.e., Model 3: $Tbabs(bvapec+bvapec)$ in Table~\ref{tab:Tab1}. The temperatures of two components, their normalisation and the $N_\text{H}$ were the only free parameters here. This fit led to a slightly worse $\chi^2_\nu$ of 1.325 for 303 DOF.  The best fit value found for $N_\text{H}$ reduced as was also found by \citet{Or2020} to the value comparable to that reported by them with Model 3.  We also found that though the temperature values in this case were nearly the same as before using $bapec$, the best fit emission measure (EM) value for the cooler component dropped significantly as was also reported by \citet{Or2020}.
Given the low resolution and insufficient signal-to-noise ratio in S1 data, varying the abundances individually to differ from the fixed values, to account for 2 days difference in observations between $Chandra$ and SXT, introduced too many variable parameters and did not lead to any useful determination of the abundances.  
We caution that the parameter values (abundances and temperatures) are only qualitative and just indicative of a possible trend as several weak lines are unresolved and can lead to wrongful estimate of continuum using global models. In addition, the atomic data base used in the analysis are continuously evolving (see the AtomDB website mentioned above). 
We believe that given the lower resolution in SXT vis-a-vis $Chandra$ HETG, and non-simultaneity of the two observations, using $bvapec$ is not necessarily warranted.  We, therefore, have limited ourselves to using two temperature $bapec$ models with variable abundances of all elements tied to together, and as shown in Table~\ref{tab:Tab1}, for all further analysis of S1 data.

We have used the {\it XMM-Newton} MOS1 and MOS2 data taken on Day 18 after the outburst and during our long S2 observations for comparison with the merged hard X-ray spectrum obtained during the S2 observations. The SXT spectrum from the entire S2 data was used so as to have a comparable signal-to-noise in the two spectra. A major dip can be seen in the MOS1 and MOS2 data in Fig.\ref{fig:smap2} which was not covered in the SXT observation. Therefore, we extracted the spectral data from MOS1 and MOS2 before and after the dip for our comparison as described in \S2.2. For the reasons mentioned earlier, all spectra viz., MOS1, MOS2 and S2 used here were for energies greater than 1.1\,keV, and are shown in the right panel of Fig.~\ref{fig:hard}.  These  three spectra were fit jointly with the same models as employed for the S1 spectrum and listed in Table~\ref{tab:Tab2}.  We first tried to fit the data with the solar abundance values by freezing the abundance parameter to unity (Model 1) and later by making it free but keeping it common for the two plasma temperature components (Model 2). We used the blue shift and velocity of 0.00289 and 632 km/s for both the low temperature and the high temperature components from \citet{Or2020}.   The spectral fits were found to be not very sensitive to these parameters within a range of few hundred km/s in the velocity parameters. We used a systematic uncertainty of 3\% in the model parameters here and in the analysis of all the S2 spectra that follow.  The values of the best fit parameters, viz., the column density, the temperatures, the emission measures of the plasma components and the abundances are given in Table~\ref{tab:Tab2}.  A significantly improved fit was obtained with sub-solar abundances with the $\chi^2_\nu$ going down from 1.72 for 488 DOF for solar abundances (Model 1) to 1.48 for 487 DOF (Model 2). The best fit abundance value is nearly half the solar value and the $N_\text{H}$ value is $\sim$35\% lower than in the case of S1. We also fitted models where the elemental abundances of Ne, Mg, Si, S and Fe were varied individually relative to the solar values but tied together to have the same value for the two temperature components (Model 3 in Table~\ref{tab:Tab2}).  The best fit parameter values obtained from a joint fit with this model to SXT S2,  MOS1 and MOS2 spectra are listed in Table~\ref{tab:Tab2}.
A further improvement is seen in the best fit ($\chi^2_\nu$ of 1.40 for 483 DOF) with the abundances of Mg, Si, S and Fe being significantly sub-solar, while the abundance of Ne is indicated to lie in the range of  1.3-3.3 times solar.
The Model 3 spectra are shown in right panel of Fig.~\ref{fig:hard} for a joint fit to the S2 + MOS1 + MOS2 spectra.

Using the models derived from fitting the non-variable hard X-ray component, we then proceeded to model the highly variable low-energy component of the spectra. We used Model 1 in Table~\ref{tab:Tab1} for the S1 data and Model 3 in Table \ref{tab:Tab2} for the S2 data, for modeling the high energy part of the spectra and added an additional component for the low energy part, as explained below.

\begin{figure*}
	\includegraphics[width=2.1\columnwidth]{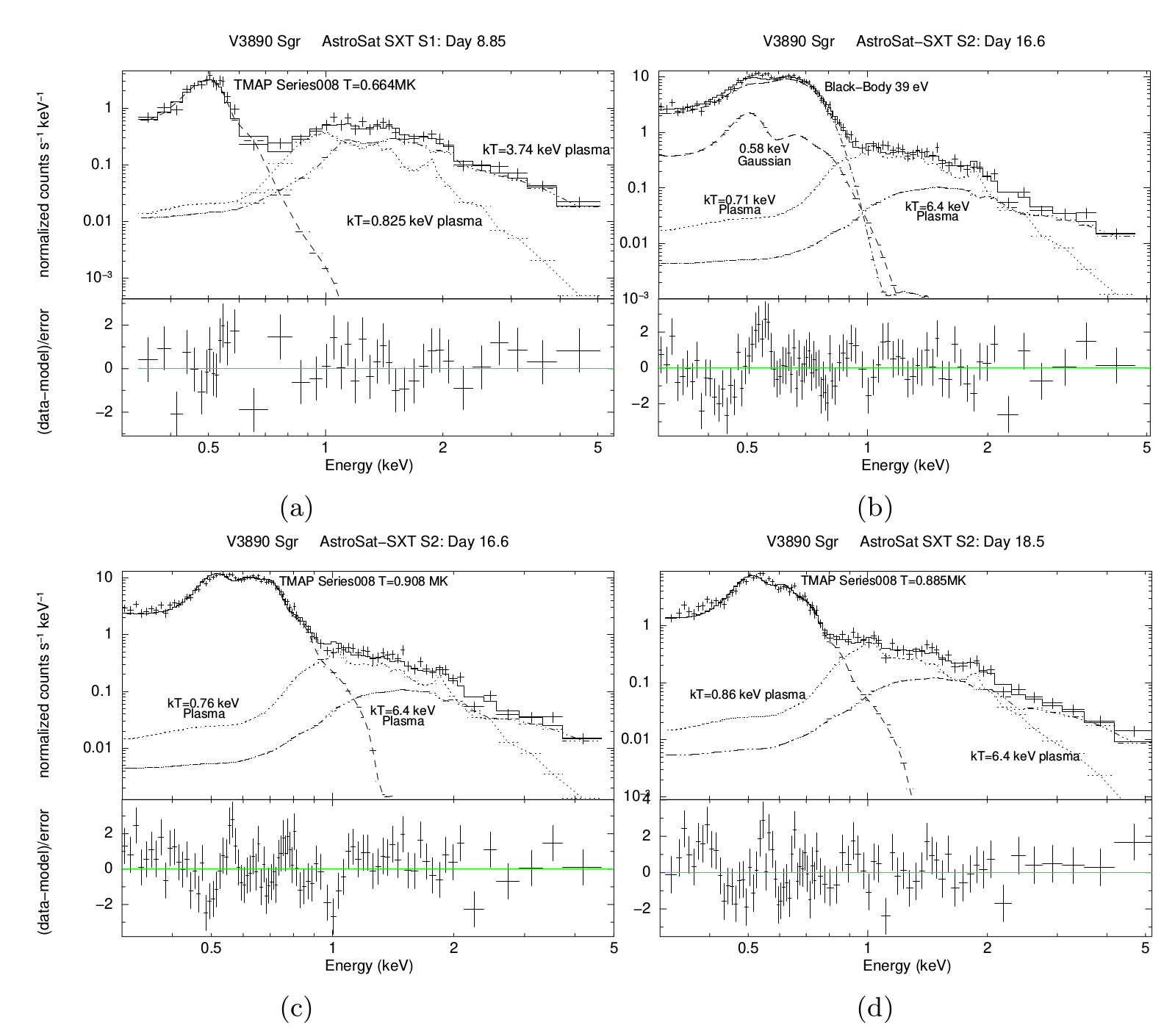}
    \caption{X-ray spectra from three epochs: one from the S1 observations on Day 8.85 (panel a), and two from S2 observations: Day 16.6 (panels b and c) and Day 18.5 (panel d). The S1 spectrum (panel a) has been fitted with Model 1 in Table 1 plus model 008 from Tübingen NLTE Model-Atmosphere Package (see text).  S2 spectra  (panels c and d) have been fitted with two temperature $bvapec$ models with non-solar abundances (Model 3 in Table 2) plus model 008 from Tübingen NLTE Model-Atmosphere Package. Spectrum in the panel (b) for Day 16.6 shows the fit with a black-body and a Gaussian line in addition to two plasma components for comparison with the fit shown in panel (c). The individual contributions of all the models are shown in each figure marked with different line styles and with the name of the model and the best fit temperatures next them. The residuals from the overall fit are shown in the bottom panels of each figure.} 
    \label{fig:fig8}
\end{figure*}

\subsubsection{Low Energy Spectral Evolution}

We modelled the spectra over the full energy range (0.3$-$7.1 keV) obtained for the duration of individual orbits (28 in case of S1 and 51 for S2) of the $AstroSat$ SXT.  As mentioned already, these spectra have useful exposure times ranging from $\sim$300s to $\sim$ 3000s (most common value being $\sim$ 1700s), and thus have low signal-to-noise ratio per energy channel. Before the appearance of the SSS phase in the spectra (for the first 8.5 days after the outburst in this case), Model 1 used in Table~\ref{tab:Tab1} was sufficient to get a good a fit, as expected. After the appearance of the SSS phase, however, an additional black-body component to represent the white dwarf atmospheric continuum was required to be added. All the spectra from the S1 observations after the day 8.5 could be well fit by model: $Tbabs(bapec+bapec+bbody)$ with a black-body temperature in the range of 40$-$50 eV and an absorber column density, $N_\text{H}$, of (1.0$\pm$0.1)$\times$10$^{22} $\text{cm}$^{-2}$ for the continuum, while keeping the parameters of two $bapec$ components fixed to the values given in Table~\ref{tab:Tab1} for Model 1 with solar abundances.  Adding just a black-body component to the spectra taken during the S2 observations, however, failed to give an acceptable $\chi^2_\nu<$2.0. Very strong residuals were observed centered in the 0.5-0.6\, keV energy band, corresponding to a hump seen in the very soft part of the spectra shown in Fig.~\ref{fig:smap2}. Addition of a very high amplitude Gaussian with best fit width of 0.09$\pm$0.01 keV brought the $\chi^2_\nu$ down to an acceptable range of 0.9 - 1.5. The Gaussian could plausibly be explained as due to unresolved lines arising from the O VII triplet (0.574,0.561,0.569 keV), N VII (0.50 keV), and/or O VIII (0.654,0.0653 keV) in an additional thin thermal plasma component of low temperature in the range of 0.17-0.27 keV.  Replacing the line component and/or the black-body with a plasma component, however, did not work to improve $\chi^2_\nu$ and the residuals between 0.5-0.6 keV persisted. Besides, the super soft emission originates from the photosphere around the white dwarf and is thus poorly represented by an optically-thin thermal plasma emission model or a black-body. We, therefore, replaced the black-body by atmospheric models for white dwarfs provided by Tübingen NLTE (non-local thermal equilibrium) Model-Atmosphere Package (TMAP){\footnote{http://astro.uni-tuebingen.de/~rauch/TMAP/TMAP.html}}. 
The models given in TMAP calculate emission from stellar atmospheres in spherical or plane-parallel geometry in hydrostatic and radiative equilibrium while allowing for departures from local thermodynamic equilibrium (LTE) for the population of atomic levels \citep{Ra1997,Ra2003,RD2003}.  These models have been used to explain the SSS sources associated with novae, e.g., V4743 Sgr \citep{Ra2010}.  These models are available as tables that can be used in the {\sc xspec} corresponding to a series of abundance values for the hot white dwarf atmospheres{\footnote{http://astro.uni-tuebingen.de/~rauch/TMAF/flux\_HHeCNONeMgSiS\_gen.html}}, and give fluxes for grids of effective temperatures ranging from 0.45\,MK to 1.05\,MK, in steps of 0.04\,MK.  The fluxes given are calculated for model series: 003, 004, 005, 0,006, 0.007, 0.008, 0.009, 0.010 and 0.011 corresponding to different sets of abundance ratios.  The abundance ratios used in different series vary principally for C and N elements, with decreasing depletion for C and decreasing overabundance of N with respect to solar values, going from model series 003 to 011. The surface gravity, $\log g$, is assumed to be 9.
The abundance ratios of other elements are nearly constant in these models. 
These abundances encompass the values expected from the processing of accreted material in CNO cycle.  We used these models for fitting the S2 data having bright SSS phase, using {\sc xspec} and found that these models could fit the double hump feature seen in spectra (Fig.~\ref{fig:smap2}) without requiring the addition of any Gaussian component while using a black-body. We tried all the model series and found that S2 data show a strong preference for TMAP model series 008.  These models also provided acceptable fits to the SSS phase observed in the S1 spectra, but showed no preference for any particular series. These models were invoked as $TBabs(bapec+bapec+ atable{SSS\_008\_00010-00060.bin\_0.002\_9.00.fits})$ in the {\sc xspec}.  Examples of fits with the TMAP model to the S1 and S2 data are shown in Figure~\ref{fig:fig8} in panels (a), (c) and (d). Panel (b) in this figure shows the fit with a black-body and a Gaussian line in addition to two plasma components for comparison with the TMAP model (panel c).

\begin{figure}
	\includegraphics[width=\columnwidth,angle=0]{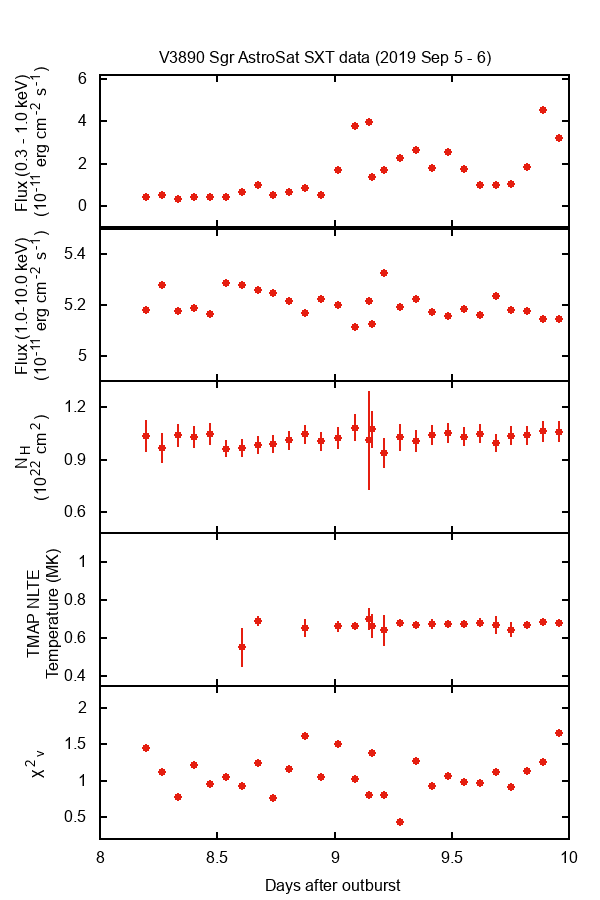}
    \caption{Spectral parameters obtained from the analysis of orbit-wise time resolved X-ray spectra during the observation S1 shown as a function of time in days after the outbursts. The errors plotted on the parameter values are with 90\% confidence. The missing data points in the panel for TMAP NLTE temperature correspond to the epochs for which the SSS phase was missing or $\sim$100 times lower than at the other epochs for which the errors have been determined (see also Fig.5)}
    \label{fig:S1_time_res}
\end{figure}

\begin{figure}
	\includegraphics[width=\columnwidth,angle=0]{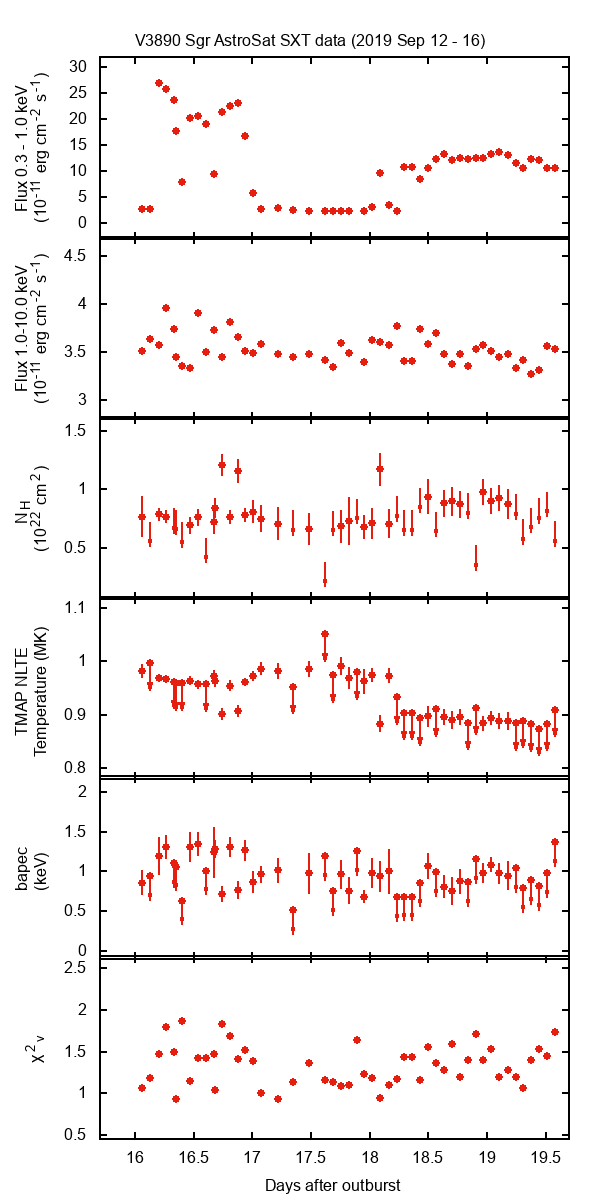} 
    \caption{Spectral parameters obtained from the analysis of the time resolved X-ray spectra during the observation S2. The errors plotted are with 90\% confidence. Upper limits are plotted for the epochs when the SSS phase has a very low flux.}
    \label{fig:S2_time_res}
\end{figure}

Finally, we fitted all S1 spectra with model: $TBabs(bapec+bapec+
atable{SSS\_008\_00010-00060.bin\_0.002\_9.00.fits})$, using the
temperatures, solar abundances, line broadening, line shift and the
normalisation of both the plasma components, $bapec$, obtained from fitting
the high energy spectra shown in Fig.~\ref{fig:hard} frozen to the values
listed in Table~\ref{tab:Tab1} (Model 1) as they represented the
non-variable hard X-ray component. The only free parameters were
$N_\text{H}$, temperature and normalisation of the TMAP model component.
The best fit values of $N_\text{H}$ and temperature of the TMAP (with their
90\% confidence error bars), and the X-ray intensity (after absorption) in
the 0.3$-$1.0\,keV and 1.0$-$10.0\,keV energy bands are plotted in
Fig.~\ref{fig:S1_time_res}.  The $N_\text{H}$ values vary in the range of
0.9$-$1.2 $\times10^{22}\text cm^{-2} $ with an average value of
1.03$\pm$0.03$\times10^{22}\text cm^{-2} $.  The contribution from a TMAP
model first becomes significant only on $\sim$day 8.6 after the outburst,
disappearing thereafter until day 8.75, reappearing briefly on day 8.8,
disappearing again before coming on permanently from day 8.85, consistent
with the data plotted in Fig.~\ref{fig:smap1}.  The normalisation of the
white dwarf atmosphere component during the act of disappearance is at
least 100 times lower than during its appearance. The temperature of the
TMAP component during the SSS phase of S1 observations is observed to have
an average value of  0.66$\pm$0.03\,MK.

Spectra from the S2 observations were also fitted with $TBabs(bapec+bapec+
atable{SSS\_008\_00010-00060.bin\_0.002\_9.00.fits})$. However, since there
is a considerable overlap at the boundary between the low temperature
$bapec$ component and the TMAP model (Fig.~\ref{fig:fig8} and
Fig.~\ref{fig:smap2}), the temperature and the normalisation of this
$bapec$ component were allowed to vary freely along with the temperature
and normalisation of the TMAP model component to get a good fit. The
$N_\text{H}$ was allowed to vary as before. Temperature of the high
temperature $bapec$ component and its normalization along with the
non-solar abundances, and line shift of the plasma components, obtained
from fitting the high energy spectra shown in Fig.~\ref{fig:hard} were
frozen to the values shown in Table~\ref{tab:Tab2} (Model 3) as they
represented the non-variable hard X-ray component.  The best fit values of
the free parameters with their 90\% confidence error bars, and X-ray
intensity (absorbed) in the 0.3$-$1.0\,keV and 1.0$-$10.0\,keV energy bands
for the S2 observation are shown plotted in Fig.~\ref{fig:S2_time_res}. We
find that even during the epochs of very low SSS flux,  the presence of
both a TMAP model and the thin plasma component was required to obtain a
good fit even though their temperatures could not be constrained.  The
column density of the line-of-sight absorber also could not be constrained
well during these epochs. These points are shown as upper limits in
Fig.~\ref{fig:S2_time_res}.  The  best fit $N_\text{H}$ values show a lot
of dispersion, ranging from 0.4$\times$10$^{22}$\text{cm$^{-2}$} to
1.2$\times$10$^{22}$\text{ cm$^{-2}$} with median value of
0.8$\times$10$^{22}$\text{cm$^{-2}$}.  The column density does not seem to
show any trend of being associated with the absence or presence of the SSS
phase. The temperatures of the low temperature plasma component are
clustered around the average value of 1$\pm$0.2\,keV, similar to the 1\,keV
value obtained from the hard X-ray spectra shown in Fig.~\ref{fig:hard}(b)
and Table~\ref{tab:Tab2}. The temperature of the TMAP model shows an
average value of 0.97$\pm$0.03\,MK, before day 18.2, declining to
0.89$\pm$0.03\,MK thereafter, with both the values being considerably
higher than the value obtained during the S1 observations.
 
\section{Discussion} \begin{enumerate} \item The two dense monitoring
		observations of the third recorded outburst of V3890\,Sgr in X-rays
		with $AstroSat$ have added useful data for understanding the
		behaviour of RNe during their eruptions.  The X-ray spectra
		recorded in these observations show an enhancement in the flux
		below 0.8 keV around day 8.57 since outburst discovery, almost
		coincident with the beginning of the SSS phase on day 8.36
		\citep{Page2019a}. As shown in Fig.\ref{fig:xrt}, the SSS phase
		peaked during days 15-20 and ended on day 26.18  \citep{Page2019b}.
		The duration of SSS phase can be as long as a decade \cite[e.g.,
		V723\,Cas,][]{Ness2008} and as short as a few days \cite[e.g., V745
		Sco,][]{Pag2015}. The mass of the WD is the main factor determining
		the duration of the SSS phase, with WDs with low mass having a
		longer duration compared to the more massive WDs \citep{Orio2008}.
		Short duration SSS phase has been observed in other recurrent novae
		such as U\,Sco, M31N\,2008-12a \citep{Hen18}, V745\,Sco and
		RS\,Oph. The SSS phase in V745 Sco was detected around 4 days since
		outburst, and lasted for $\sim$2\,days only, making it the earliest
		switch-on and shortest duration SSS ever detected \citep{Pag2015}.
		The early switch-on is attributed to a combination of low ejected
		mass with high velocity, high effective temperature and a high mass
		WD (>1.3\,M$_\odot$) \citep{Pag2015}. RS\,Oph, on the other hand
		showed the SSS onset on day 26 during the 2006 outburst, with the
		SSS phase lasting for $\sim$60\,days \citep{Bode2006, Osb2011}. A
		high mass WD is inferred for RS\,Oph also \citep{Osb2011}. The
		early onset of the SSS phase in V3890\,Sgr indicates a high mass WD
		in this RN too. Adopting the relation between the SSS turn-on and
		ejected mass by \cite{Sho08}, $M_{\text{ej}}\sim 2\times10^{-6}$
		M$_\odot$ is estimated. An accretion rate of $\sim$2$\times$
		10$^{-7}$ M$_\odot$\,yr$^{-1}$ is estimated using the relations in
		\citet{Pag2015} and \cite{Hen14}, and by adopting super-Eddington
		luminosity as suggested by \cite{Gon1992} for the 1990 outburst.
		These estimates indicate that not all the accreted material is
		ejected during the nova outburst. Theoretical models show that WDs
		of mass 1.35 M$_\odot$ increase in mass irrespective of the
		accretion rate and could eventually reach the Chandrasekhar limit
		\citep{Star2020}.

\medskip \item A detailed spectral analysis of our two long observations
(day 8.1-9.9, and day 15.9-19.6 after outburst) has revealed at least two
plasma components dominating the emission at energies above 1\,keV,  while
an additional soft component dominates at lower energies. The average X-ray
spectrum above 1\,keV taken during our first set of observations (S1) is
consistent with $Chandra$ high-resolution X-ray spectra taken 2 days before
the start of our observations. The temperatures of $\sim$0.8\,keV and
$\sim$4.0\,keV for the two plasma components and the absorbing column
density value of $\sim$10$^{22}$ \text{cm$^{-2}$} derived from global fits
are similar to the values reported based on $Chandra$ observations based on
similar analysis. The EM values of the plasma components here (see
Table~\ref{tab:Tab1}) are only slightly lower than those given in Table 2
of \cite{Or2020}. In our case, the EM values are in the range
(0.6$-$0.7)$\times$10$^{58}$ \text{cm$^{-3}$} for the cool (kT=0.8\,keV)
component and $\sim$ 1.1$\times$10$^{58}$ \text{cm$^{-3}$} the hot
(kT$\sim$4.0\,keV) plasma component (Table~\ref{tab:Tab1}). When the
variable abundance model $bvapec$ is used the EM for the cool component
goes down considerably to 3.5$\times$10$^{55}$ \text{cm$^{-3}$}, similar to
the trend seen in Table 2 of \cite{Or2020}. These values, though model
dependent, can be used to get a rough idea of the electron density in the
plasma if the size of the region responsible for the plasma components is
known. If the plasma components are assumed to be coming from a region of
radius 1\,AU, then the electron density in the plasma would range from
$7-9\times 10^{8}$\text{cm$^{-3}$}. This is somewhat lower than the typical
values derived from optical spectra of other novae at early phases. For
example, \citet{Neff78} estimated a value
1.7$\times$10$^{10}$\text{cm$^{-3}$} on the 9th day after the outburst for
V1500 Cyg. Further the models by \citet{Moo12}, \citet{Orl09}, and
\citet{Orl17} indicate post-shock electron density in the range of 10$^{9}
- 10^{10}$\text{cm$^{-3}$} for the post-shock material in symbiotic
recurrent novae assuming a spherically symmetric expansion of the plasma.
We will mention, however, that in the $Chandra$ observation on day 6,
\citet{Or2020} found spectral diagnostics suggesting the possibility of a
very dense, localised and probably clumpy emission region, occupying only a
tiny part of the 1AU radius sphere. In this case, the X-ray spectrum at
that epoch would have probably been emitted very close to the red giant,
with electron density as high as 10$^{13}$\text{cm$^{-3}$}. 

\medskip \item We have also analysed MOS1 and MOS2 data obtained from {\it
XMM-Newton} observations on an epoch during our second set of observations,
S2, and carried out a joint spectral fit to the high energy (>1.1\,keV)
MOS1 + MOS2 + SXT data in the same energy band. The SXT data were, however,
merged for the entire duration of the S2 observations to have comparable
statistics in this steady high energy part of the spectrum. We find that
the fluxes in this energy range of 1.1$-$7.1 keV agree to within 20\% for
the SXT and MOS1 (MOS2 data taken in a different mode show a much lower
flux), while all the spectral parameters (see Table~\ref{tab:Tab2}) are the
same for all three spectra. A comparison with the best fit parameters
derived from the S1 observations shows that: the temperatures of the thin
equilibrium plasma components are somewhat higher,  the column density is
about 30 percent lower and the elemental abundances are significantly lower
(about half) with respect to solar. The best fit $N_\text{H}$ values,
however, show a lot of dispersion for the duration of the S2 observation
with median value of 0.8$\times$10$^{22}$\text{cm$^{-2}$}. The trend for
decreasing column density has also been reported by \cite{Page2019b} who
saw a decrease of column density from $3.4\times 10^{22}$\text{cm$^{-2}$} on
day 8.1 to $0.17^{+0.5}_{-0.6} \times 10^{22}$\text{cm$^{-2}$} on day 28.8
after the outburst. The exact values, however, are dependent on the
abundance model and abundance table used in the analysis. Decrease in the
column density was also detected in both RS Oph and V745 Sco, with
($N_{\text{H}} \propto t^{-\alpha}$) where $\alpha$ was found to be $\sim$0.5
in RS Oph, while it was somewhat steeper with $\alpha=0.76$ in V745 Sco
\citep{Bode2006,Pag2015}.  This could be the result of complex mixing
processes of nova ejecta and red giant wind as the nova evolves.  In fact
model calculations by \citet{Orl09} indicate that different regions of
plasma at different temperatures could also have a different mixture of
elements, and quite different abundances.

\medskip \item The sum of the two plasma components is steady in two SXT
observations while the source intensity varies erratically at energies
below 1 keV.  A soft component dominates the spectra from $\sim$8.57\,day
onwards, and is responsible for all the variations in the intensity
observed after that day. The soft component below <1\,keV could be modelled
either using TMAP models or a black-body for both the S2 spectra and S1
spectra (after the appearance of the SSS phase). The addition of a
black-body alone was not sufficient to fit the spectra from S2
observations, however, and required the addition of a narrow Gaussian
component in the energy range of 0.5$-$0.65\,keV. The Gaussian component
could likely originate due to transitions of the O VII (0.574,0.561,0.569
keV), N VII (0.50 keV) from a plasma at a temperature of 0.172 keV, and O
VIII (0.654,0.653 keV) from a plasma of temperature 0.272 keV.  The TMAP
models, however, provided good fit to S2 spectra without needing any
additional Gaussian component, and also fitted the S1 spectra quite well.
We, therefore, studied the evolution of the soft component using only the
TMAP models.  The best fit TMAP models show a preference for model series
008 which shows a lesser depletion of C and N in CNO burning as compared to
the depletion in model 003 that was used by \citet{Osb2011} to study the
SSS phase of RS Oph. These abundance values also impact the determination
of the mass of the WD as shown by \citet{Star2020}. The temperature of the
TMAP models shows the presence of a very hot photosphere with temperature
increasing from 0.68\,MK to an average of 0.94\,MK from S1 observations to
S2 observations, with an indication of slight cooling from a peak of
1.0\,MK to <0.9\,MK after day 18.2. The temperature of the low temperature
plasma component also seems to have increased from 0.83\,keV in S1 data to
$\sim$1.0\,keV in S2 data.  If the low temperature components disappear
during days subsequent to the disappearance of the black-body as reported
in Swift observations nearly 30 days after the eruption, then it would
indicate low cooling time due to high density in the low temperature
components. 

\medskip \item  The complete disappearance of SSS phase seen as very low
flux lasting for a day (16.8 - 17.8\,d after the outburst) during the S2
observations, and observed as a sharp dip a little later with {\it
XMM-Newton} could be the result of a sudden new outflow of matter that is
completely opaque to X-rays but does not fill the space, thereby reducing
the flux but not the spectral shape or $N_{\text{H}}$ or patchy absorption
by clumpy ejecta.
It does not appear to be associated with increased overall absorption,
although the determination of the $N_{\text{H}}$ is complicated by low flux
and poor signal-to-noise ratio during some of these epochs. In fact the
$N_{\text{H}}$ values show a wide range after the addition of TMAP models,
as compared with a low value obtained from the average high energy spectra
from S2 observations, thus indicating the complexity of ejecta, chemical
composition and mixing with accretion from symbiotic giant star.

 The reasons for the rapid variability on time scales of several minutes to hours observed in the intensity of very soft X-rays (0.3 - 1.0 \,keV) are not clear, however.

\medskip
    \item 
The complex interplay of black-body like emission and coronal lines versus TMAP models requires monitoring with high spectral resolution. For example, coronal lines have been reported from high resolution $Chandra$ observations of the super soft source Cal87 in the Large Magellanic Cloud \citep{Ness2012usco,Ness2013}.
The presence of coronal line emission due to OVI has been seen in the optical spectra from day 12 onwards by \citet{PaM2019}, along with the other coronal lines due to [Fe XIV], [Fe VII], [Fe X] and [Fe XI] during the active SSS phase.  Our optical data (M. Pavana et al. in preparation) also indicate the optical coronal emission lines originate in a region very close to the red giant.
\end{enumerate}

\section{Conclusions}
 The two long extended observations with $AstroSat$ cover the nova
evolution with the densest possible monitoring from a low-Earth orbit. It demonstrates the fast evolution of SSS emission including a rapid first appearance on day 8.57 after the outburst. 
The subsequent evolution remains highly variable
demonstrating the importance of long-term high-cadence monitoring. 
A complete vanishing of the supersoft emission during days 8.6$-$8.9 followed by another extremely low flux state lasting for a day (day 16.8-17.8) has also been detected. The abruptness with which SSS emission brightens and fades (or disappears) poses a challenge to evolutionary models of the ejecta, a discussion on which is beyond the scope of this work.
Long duration continuous observations with higher resolution instruments of future outbursts in such novae are needed to get a better understanding of the evolution of such systems. 

\section*{Acknowledgements}
We are thankful to Kim Page of the University of Leicester, UK, for providing us with data from observations with $Swift$ for our Fig. 1. We thank the Indian Space Research Organisation for scheduling the observations within a short period of time and the Indian Space Science Data Centre (ISSDC) for making the data available. This work has been performed utilizing the calibration data-bases and auxiliary analysis tools developed, maintained and distributed by $AstroSat$-SXT team with members from various institutions in India and abroad and the  SXT Payload Operation Center (POC) at the TIFR, Mumbai for the pipeline reduction.The work has also made use of software, and/or web tools obtained from NASA's High Energy Astrophysics Science Archive Research Center (HEASARC), a service of the Goddard Space Flight Center and the Smithsonian Astrophysical Observatory.  We thank the {\it XMM-Newton} team for carrying out the observations. {\it XMM-Newton} is an ESA science mission with instruments and contributions directly funded by ESA Member States and NASA. We thank an anonymous referee for the comments that substantially improved the presentation of our analysis and results.\\

\section*{Data Availability:} Data from observations used in this paper are
publicly available at the $AstroSat$ archives of ISRO maintained by the
ISSDC{\footnote{
$https:\/\/astrobrowse.issdc.gov.in\/astro\_archive\/archive\/Home.jsp$}}. Data
from {\it XMM-Newton} observations are also public and can be retrieved via
XMM-Newton Science Archive (XSA)
{\footnote{http://nxsa.esac.esa.int/nxsa-web/\#search}}.



\bibliographystyle{mnras}


\bsp	
\label{lastpage}
\end{document}